\documentclass[12pt]{article}
\usepackage{graphicx,epsfig}

\newcommand{\ee}{\end{equation}}
\newcommand{\be}{\begin{equation}}
\def\ba{\begin{eqnarray}}  \def\ea{\end{eqnarray}}
  \def\hy{hyperbolic } \def\hyp{hyperbola}
   \def\r{\rho}  
  \def\Th{\Theta} \def\th{\theta} 
 \def\f{\phi}   
  \def\a{\alpha} \def\b{\beta} \def\g{\gamma}
\def\c{\gamma}

\def\ra{\rightarrow} \def\Ra{\Rightarrow} \def\2{{1\over2}}
\def\lra{\leftrightarrow}\def\Lra{\Leftrightarrow}

\begin{document}
\title{Hyperbolic trigonometry in two-dimensional space-time geometry}
\author{\sc F. Catoni, R. Cannata, V. Catoni, P. Zampetti
\vspace{4mm} \\ \it ENEA; Centro Ricerche Casaccia; \vspace{1mm} \\
\it Via Anguillarese, 301;
00060  S.Maria di Galeria; Roma; Italy}
%{\it ENEA, Centro Ricerche Casaccia-Via Anguillarese 301, 00060
%S.Maria di Galeria \\ Roma, Italy} \\ [4mm]
\date{January 22, 2003}
\maketitle
{\bfseries Summary.-}
By analogy with complex numbers, a system of hyperbolic numbers 
can be introduced in the same way:
$\{z=x+h\,y;\;h^2=1\;\;x,\,y\in {\bf R}\}$.
As complex numbers are linked to the Euclidean geometry, so this
system of numbers is linked to the pseudo-Euclidean plane geometry (space-time
geometry).\\
In this paper we will show how this system of numbers allows, by means of a
Cartesian representation, an operative 
definition  of \hy  functions using the invariance respect 
to special relativity Lorentz group. From this definition, by using elementary 
mathematics and an Euclidean approach, it is straightforward to formalise the pseudo-Euclidean
trigonometry in the Cartesian plane with the same coherence as the 
Euclidean trigonometry. \\[2mm]
PACS 03 30 - Special Relativity 
\section{Introduction}
Complex numbers are strictly related to the Euclidean geometry: indeed their
invariant (the module) is the same as the Pythagoric distance (Euclidean
invariant) and their unimodular multiplicative group is the Euclidean
rotation group. It is well known that these properties allow to use 
complex numbers for representing plane vectors. \\
In the same way hyperbolic numbers, an extension of complex numbers
\cite{La, Ya2} defined as $$\{z=x+h\,y; \,\,h^2=1\,\,\,x,y \in {\bf R}\},$$
are strictly related to space-time geometry \cite{Ya2, cz, ca}. 
Indeed their square module given by\footnote{We call $\tilde z=x-h\,y$ the \hy
conjugate of $z$ as for complex numbers} $|z|^2=z\tilde z\equiv x^2-y^2$
is the Lorentz invariant of two dimensional special relativity, 
and their unimodular multiplicative group is the special relativity Lorentz 
group \cite{Ya2}. These relations have been used to extend special relativity 
\cite{Fj}. Moreover by using the functions of 
hyperbolic variable the two-dimensional special relativity has been generalised
\cite{cz}. These applications make the \hy  numbers relevant for physics 
and stimulate the application of hyperbolic numbers, just as complex numbers 
are applied to the Euclidean plane geometry \cite{Ya2}. In this paper we 
present the basic concepts of space-time trigonometry which we derive from the
remark that hyperbolic (complex) numbers allow to introduce two invariant
quantities in respect to Lorentz (Euclid) group. The first invariant is
the scalar product, recently considered by Fjelstad and Gal \cite{Fj2} as the 
basis of their paper. The second one is equivalent to the module of the 
vector product (i. e., an area).
These two invariant quantities allow to define, in a Cartesian 
representation, the trigonometric hyperbolic functions. These functions
are defined in the whole hyperbolic plane and allow to solve triangles having
sides of whatever direction\footnote{We shall see that only 
side directions parallel to axes bisectors have to be excluded.}.   
Then the space-time trigonometry and geometry can be formalised in a 
self-consistent axiomatic-deductive 
way, that can be considered equivalent to Euclidean geometry construction. 
More precisely we start from the experimental axiom that the Lorentz
transformations do hold and we look for the geometrical ``deductions".

It is well known that all the
theorems of the Euclidean trigonometry are obtained through
elementary geometry observations. In fact once we define in a Cartesian plane
the trigonometric functions as a direct consequence of Euclid rotation 
group (as shown in the appendix \ref{inve}), all the trigonometry theorems  
follow just as mathematical identities. In our work, since we do not have 
the same intuitive vision for the pseudo-Euclidean geometry, we make up 
for this lack of evidence by using an algebraic approach and just checking 
the validity of the \hy trigonometry theorems by elementary mathematics.
This description can be considered similar to the Euclidean representations 
of the non-Euclidean geometries obtained in XIX century, i.e., to the E. 
Beltrami's interpretation, on constant
curvature surfaces, of the non-Euclidean geometries \cite{Bel}.

The paper is organised in the following way: in section 2 some properties 
of \hy numbers are briefly resumed. In section 3  the 
trigonometric \hy functions are derived as a consequence of invariance 
with respect to Lorentz group. In section 4 the trigonometry in the 
pseudo-Euclidean plane is formalised. Three appendixes are added 
to clarify some specific aspects.

\section{Hyperbolic numbers}
\subsection{Basic concepts} 
Here we briefly resume some fundamental properties of \hy  numbers.
This number system has been introduced by S. Lie \cite{Lee} as a two
dimensional example of the more general class of the commutative 
hypercomplex number systems\footnote{ \label{hy}
Hypercomplex numbers \cite{Lee, Ca} are defined by the expression:
${\displaystyle x=\sum^{N-1}_{\a =0} e_{\a}x^{\a}}$ where $x^{\a}\in${\bf R} 
are called {\it components} and $e_{\a}\notin${\bf R} units or {\it versors},
as in vector algebra. This expression defines a hypercomplex 
number if the versors multiplication
rule is given by a linear combination of versors: ${\displaystyle 
e_{\a}e_{\b}=\sum^{N-1}_{\g =0} C_{\a \b}^{\g}e_{\g}}$    
where $C_{\a \b}^{\g}= C_{\b \a}^{\g}$ are real constants, called 
{\it structure constants}, that define the characteristics of the system \cite{Ca}. \\ 
The versor product defines also the product of hypercomplex numbers. 
This product definition makes the difference between vector algebra and 
hypercomplex systems and allows to relate the hypercomplex numbers to groups. 
In fact the vector product is not, in general, a vector while the product of
hypercomplex numbers is still a hypercomplex number; the same for the division,
that for vectors does not exist while for hypercomplex numbers, in general,
does exist.}. In more recent years the \hy numbers have been considered by
B. Chabat \cite{La} for studying ultrasonic phenomena, and by P. Fjelstad 
\cite{Fj} who called them ``Perplex  numbers", for extending the special 
relativity to represent the superluminal phenomena. 

Let us now introduce a \hy  plane by analogy with the Gauss-Argand plane of the
complex variable. In this plane we associate the points $P\equiv(x,\,\,y)$ to 
\hy numbers $z=x+h\,y$.  If we represent these numbers on a Cartesian 
plane, in this plane the square distance of the point $P$ from the origin of 
the coordinate axes is defined as 
\be
D=z\tilde z\equiv x^2-y^2. \label{disq}
\ee 
The definition of distance (metric element) is equivalent to introduce the
bilinear form of the {\it scalar product}. The scalar product and the properties
of hypercomplex numbers allow to state suitable axioms \cite[p. 245]{Ya2} 
and to give to the pseudo-Euclidean plane the structure of a vector space.       

Let us consider the multiplicative inverse of $z$ that, if it exists, is given by:
$1/z\equiv \tilde z/z\tilde z.$ This implies that $z$ does not have an 
inverse when $z\tilde z\equiv x^2-y^2=0$, i.e., when $y=\pm x$, or alternatively when 
$z=x\pm h\,x$. These two straight-lines in the \hy plane, whose 
elements have no inverses, divide the \hy  plane in four sectors that 
we shall call {\it Right sector (Rs), Up sector (Us), Left sector (Ls),} and 
{\it Down sector (Ds)}. This 
property is the same as that of the special relativity representative plane
and this correspondence
gives a physical meaning (space-time interval) to the definition of distance.
Let us now consider the quantity $x^2-y^2$, which is positive in the $Rs, Ls\;
(|x|>|y|)$ sectors, and negative in the $Us, Ds\;(|x|<|y|)$ sectors. This 
quantity, as known from special relativity, must have its sign and appear 
in this quadratic form. In the case we should use the 
linear form  (the module of \hy  numbers), we follow 
the definition of I. M. Yaglom \cite[pag. 180]{Ya2} and B. Chabat 
\cite[p. 51]{La}, \cite[p. 72]{Ch}:
\be
\r =\sqrt{|z\,\tilde z|}\equiv\sqrt{|D|} \label{mod}
\ee
where $|D|$ is the absolute value of the square distance.
\subsection{Hyperbolic  exponential function and \hy polar transformation}
The \hy  exponential function in the pseudo-Euclidean geometry plays the same
important role as the complex exponential function in the Euclidean geometry.
By comparing absolutely convergent series it can be written \cite{La, Fj}:\\
if $|x|>|y|$:
\be
x+h\,y=\mbox{sign}(x)\exp[\r'+h\,\th]\equiv
\mbox{sign}(x)\exp[\r'](\cosh \th+h\,\sinh \th)     \label{exp}
\ee
if $|x|<|y|$:
\be
x+h\,y=\mbox{sign}(y)\exp[\r'+h\,\th]\equiv
\mbox{sign}(y)\exp[\r'](\sinh \th + h\,\cosh \th)     \label{expS}.
\ee

The exponential function allows to introduce the {\sf \hy polar
transformation}. \\
Following \cite{La, Fj} we define the {\it  radial coordinate} as: \\
$\exp[\r']\Ra \r=\sqrt{|x^2-y^2|}$ \\ and {\it the angular coordinate} as:\\
$\mbox{ for } |x|>|y|:\;\; \th=\tanh^{-1}(y/x) ; \;\;\;\;\mbox{ for } |x|<|y|: \;\;
\th=\tanh^{-1}(x/y) $. \\
Then for $|x|>|y|,\,\,x>0$ (i.e., $x,\,y\in Rs$), the \hy polar  transformation is
defined as:
\be
x+h\,y \Lra \r \exp [h\th] \equiv  \r(\cosh\th+h\sinh\th). \label{expr}
\ee
Eq. (\ref{expr}) represents the map for $x,\,y\in Rs$: the map of the
complete $x,\,y$ plane is reported in tab. \ref{hip}.

\begin{table}[h]
\caption{Map of the complete $x,\,y$ plane by \hy polar
transformation} \label{hip}
\begin{center}
\begin{tabular}{|c|c|c|c|} \hline
{\it Right sector }&{\it Left sector }&{\it Up sector }&{\it
Down sector} \\
{\it (Rs)}&{\it (Ls) }&{\it  (Us)}&{\it
 (Ds)} \\[1mm]
\multicolumn{2}{|c|}{$|x|>|y|$} & \multicolumn{2}{c|}{$|x|<|y|$} \\ \hline
$z=\r\exp[h\,\th]$ & $z=-\r\exp[h\,\th]$ & $z=h\,\r\exp[h\,\th]$ & $z=-h\,\r\exp[h\,\th]$ \\ \hline
$x=\r\cosh \th$ & $x=-\r\cosh \th$ & $x=\r\sinh \th$ & $x=-\r\sinh \th$ \\
$y=\r\sinh\th$ & $y=-\r\sinh\th$ & $y=\r\cosh\th$ & $y=-\r\cosh\th$ \\ \hline
\end{tabular}
\end{center}
\end{table}
\begin{figure}[h]
\begin{center}
\epsfysize=3in\epsfbox{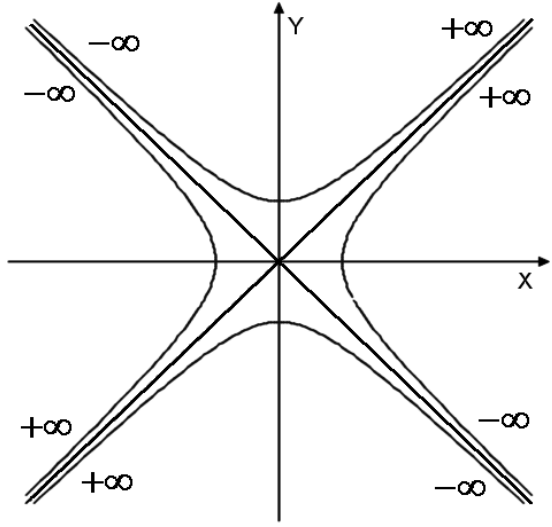}
\caption{For $\r=1$ the $x,\,y$ in tab. (\ref{hip}) represent the four
arms of equilateral hyperbolas $|x^2-y^2|=1$. Here we indicate how each
arm is traversed as parameter $\th$ goes from $-\infty$ to $+\infty$.}
\label{iperbol}
\end{center}
\end{figure}
\section{Basis of \hy trigonometry}
\subsection{Hyperbolic  rotation invariants in the pseudo-Euclidean plane geometry} \label{inv}
By analogy with the Euclidean trigonometry approach summarised in appendix 
(\ref{inve}), we can say that the {\it pseudo-Euclidean plane geometry}
studies the properties that are invariant by Lorentz 
transformations (Lorentz-Poincar\`e group of special relativity) corresponding 
 to \hy  rotation \cite{Ya2, cz}. We show afterwards, how these 
 properties can be represented by hyperbolic numbers.

Let us define in the \hy plane a \hy  vector from the origin to the 
point $P\equiv(x,\,y)$, as $v=x+h\,y$: 
a \hy  rotation of an angle $\th$ transforms this vector in a new vector 
$v'\equiv v\exp [h\,\th]$. Therefore  we can readily verify that the 
quantity:
\be
|v'|^2 \equiv v'\tilde v'= {v\exp [h\,\th]\tilde v\exp
[-h\,\th]} \equiv |v|^2 \label{modh}
\ee
is invariant by \hy  rotation. In a similar way we can find two 
invariants related to any couple of vectors.
Let us consider two vectors $v_1=x_1+h\,y_1$ and $v_2=x_2+h\,y_2$: we have 
that {\it the real and the \hy   parts of the product $v_2 \tilde v_1$
are invariant by \hy   rotation.} \\
In fact $v_2'\tilde v_1'= v_2\exp [h\a]\tilde v_1\exp [-h\a]\equiv
v_2\,\tilde{v}_1$. These two invariants allow an operative definition of the
\hy trigonometric functions. To show this let us suppose that $|x_1|>|y_1|,\;\,
|x_2|>|y_2|$ and $x_1,\,x_2>0$, and let us represent the two vectors in \hy  polar
form: $v_1=\r_1\exp [h\,\th_1],\,v_2=\r_2\exp [h\,\th_2]$. Consequently we have
\be
  v_2\bar v_1\equiv \r_1\,\r_2 \exp[h(\th_2-\th_1)]
  \equiv \r_1\,\r_2 [\cosh (\th_2-\th_1)+h\sinh (\th_2-\th_1)]. \label{invpol}
\ee
As shown in appendix (\ref{inve}) for the Euclidean plane, the real part of
the vector product represents the scalar product, while the imaginary part 
represents the area of the parallelogram defined by the two vectors. 
In the pseudo-Euclidean plane, as we know from differential geometry \cite{ei},
the real part is still the scalar product; as far as the hyperbolic part is 
concerned, we will see in section (\ref{trigpe}) that it can be considered as 
a {\it pseudo-Euclidean area}. \\
In Cartesian coordinates we have: 
\be
v_2\,\tilde v_1=(x_2+h\,y_2)(x_1-h\,y_1)\equiv x_1\,x_2-y_1\,y_2
+h(x_1\,y_2-x_2\,y_1).\label{invcar} 
\ee
By using eqs. (\ref{invpol}) and (\ref{invcar}) we obtain:
\ba
\cosh(\th_2-\th_1)=\frac {x_1\,x_2-y_1\,y_2}{\r_1\,\r_2}\equiv\frac
{x_1\,x_2-y_1\,y_2}{
\sqrt{|(x^2_2-y^2_2)||(x^2_1-y^2_1)|}} \label{cos} \\
\sinh(\th_2-\th_1)=\frac{x_1\,y_2-x_2\,y_1}{\r_1\,\r_2}
\equiv \frac{x_1\,y_2-x_2\,y_1}{ \sqrt{|(x^2_2-y^2_2)||(x^2_1-y^2_1)|}}
\label{sen}.
\ea
If we put  $v_1\equiv (1;\,0)$ and $\th_2,\,x_2,\,y_2 \ra \th,\,x,\,y$ then 
eqs. (\ref{cos}), (\ref{sen}) become:
\be
\cosh \th=\frac{x}{\sqrt{|x^2-y^2|}}; \;\;\;\;\;\;\;\; \;\;\;\;\;\;\;\;
\sinh \th=\frac{y}{ \sqrt{|x^2-y^2|}}     \label{sen1}.
\ee
The classic \hy  functions are defined for $x,\,y \in Rs$. We observe 
that expressions in eq. (\ref{sen1}) are valid for 
$\{x,\,y\in {\bf R}\,|\; x\neq\pm y\}$ so they allow  to extend the \hy  functions 
in the complete $x,\,y$ plane. This extension is the same as that already 
proposed in \cite{Fj, Fj2}, that we recall in appendix (\ref{estip}). In the 
following of this paper we will denote with
$\cosh_e ,\,\sinh_e$ these {\sf extended \hy  
functions}. In tab. \ref{tab2} the relations between $\cosh_e ,\,\sinh_e$ 
and traditional \hy  functions are reported. \\
By this extension the \hy polar transformation, eq. (\ref{expr}), is given by
\be
x+h\,y\Ra \r(\cosh_e\th + h\sinh_e\th),
\ee
from which, for $\r=1$, we obtain the {\it extended \hy Euler formula} 
\cite{Fj}:
\be
\exp_e [ h\th] =\cosh_e\th + h\sinh_e\th. \label{cosi}
\ee
\begin{table}[h]
\caption{Relations between functions $\cosh_e,\,\sinh_e$ obtained from eq.
(\ref{sen1}) and classic \hy  functions. The \hy angle $\th$ in the last four
columns is calculated referring to semi-axes $x,\,-x,\,y,\,-y$, respectively.} \label{tab2}
\begin{center}
\begin{tabular}{c|c|c|c|c|} \cline{2-5}
& \multicolumn{2}{c|}{$|x|>|y|$}&
\multicolumn{2}{c|}{$|x|<|y|$} \\ \cline{2-5}
&$(Rs),\,x>0$&$(Ls),\,x<0$&$(Us),\,y>0$&$(Ds),\,y<0$ \\ \hline
\multicolumn{1}{|c|}{$\cosh_e\th=$}&$\cosh \th$&$-\cosh \th$& $\sinh \th$&$-\sinh \th$ \\
\multicolumn{1}{|c|}{$\sinh_e\th=$}&$\sinh\th$&$-\sinh\th$&
$\cosh\th$&$-\cosh\th$ \\ \hline
\end{tabular}
\end{center}
\end{table}

\begin{figure}[h]
\begin{center}
\mbox{\epsfysize=3.74in\epsfbox{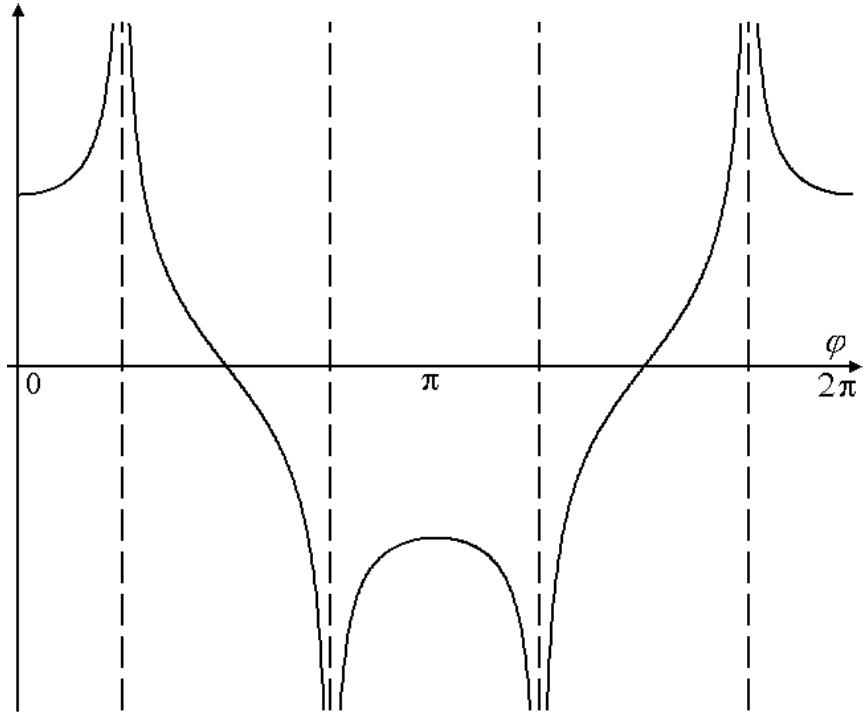}}\\[5mm]
\caption{The function $\cosh_e\th=\frac{\cos\f}{\sqrt{|\cos 2\f|}}$
for $0\leq\f<2\pi$. The dotted vertical lines represent the values for which
$\cos 2\f=0\Ra x=\pm y$. Because $\sin\f=\cos(\f-\pi/2)$, the function
 $\sinh_e$ has the same behaviour of $\cosh_e$ allowing for a shift of $\pi/2$.}
\label{funzioni}
\end{center}
\end{figure}

From Tab. \ref{tab2} or equations (\ref{sen1}) it follows that:
\ba
\mbox{ for }|x|>|y| & \Ra & \cosh_e^2-\sinh_e^2=1; \nonumber \\
\mbox{ for }|x|<|y| & \Ra & \cosh_e^2-\sinh_e^2=-1 \label{quad}
\ea
The complete representation of the extended \hy functions can be obtained
by giving to $x,\,y$ all the values on the circle $x=\cos\f,\,y=\sin\f$
for $0\leq \f<2\pi$: in this way eqs. (\ref{sen1}) become:
\be
\cosh_e\th=\frac{\cos\f}{\sqrt{|\cos 2\f|}}\equiv\frac{1}{\sqrt{|1-\tan^2\f|}};\;\;\;\; 
\sinh_e\th=\frac{\sin\f}{\sqrt{|\cos 2\f|}}\equiv\frac{\tan\f}{\sqrt{|1-\tan^2\f|}}.\label{cos2}
\ee
These equations represent a bijective mapping between the points on unit
circle (specified by $\f$) and the points on unit hyperbolas (specified by $\th$).
From a geometrical point of view  eq. (\ref{cos2}) represent the projection,
from the coordinate axes origin, of the unit circle on the unit hyperbolas.
A graph of the function $\cosh_e$ is reported in fig. \ref{funzioni}. \\
The fact that the extended \hy functions can be represented in terms of
just one expression (eqs. \ref{sen1}) allows a direct application of 
these functions for the solution of triangles with sides in any direction,
except the directions parallel to axes bisectors.
\subsection{The pseudo-Euclidean Cartesian plane}
\begin{figure}[h]
\begin{center}
\setlength{\unitlength}{.15pt}
\begin{picture}(1000,1000)(0,0)
\put(50,50){\rule[-0.200pt]{125pt}{0.400pt}}
\put(50,50){\rule[-0.200pt]{0.400pt}{125pt}}
\put(50,881){\rule[-0.200pt]{125pt}{0.400pt}}
\put(881,50){\rule[-0.200pt]{0.400pt}{125pt}}
\put(50,300){\vector(1,0){800}} \put(300,50){\vector(0,1){800}}
\multiput(52,152)(150,150){5}{\line(1,1){100}}
\multiput(52,850)(150,-150){5}{\line(1,-1){100}}
\put(830, 380){\makebox(0,0)[r]{$x$}} \put(340, 820){\makebox(0,0)[r]{$y$}}
\thicklines
\put(400,500){\line(1,2){170}}\put(400,500){\line(-1,-2){220}}
\put(400,500){\line(2,1){430}}\put(400,500){\line(-2,-1){340}}
\end{picture}
\caption{Two pseudo-orthogonal straight-lines} \label{rettepo}
\end{center}
\end{figure}
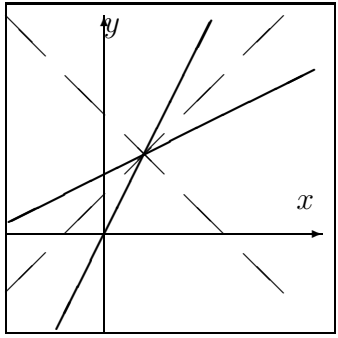

Now we show how some classical definitions and properties of the Euclidean
plane must be restated for the pseudo-Euclidean plane. \\
\noindent{\it $\bullet$ Definitions.}\\
Given two points $P_j\equiv z_j\equiv (x_j,\,y_j) ,\;P_k\equiv z_k\equiv 
(x_k,\,y_k)$ we define the {\it "square distance"} between them by extending 
eq. (\ref{disq}):
\be
D_{j,\,k}=(z_j-z_k)(\tilde z_j-\tilde z_k). 
\label{disq1}
\ee
{\it As a general rule we indicate the square segment lengths by capital 
letters, and by the same small letters the square root of their absolute value.}
\be 
d_{j,\,k}=\sqrt{|D_{j,\,k}|}. \label{disq2}
\ee
Following \cite[pag. 179]{Ya2}, a segment or
line is said to be of the {\it first (second) kind} if it is parallel to a
line through the origin located in the sectors containing the axis 
{\it Ox (Oy)}. Then the segment $\overline{P_j P}_k$ is of the first (second)
kind if $D_{j,\,k}>0$ ($D_{j,\,k}<0$)\footnote{If we give to $x$ the physical
meaning of a time variable, these lines correspond to the {\it spacelike}
({\it timelike}) lines, as defined in special relativity \cite{Na}, 
\cite[pag. 251]{Ya2}}. \\
\noindent{\it $\bullet$ Straight-lines equations.}\\
In the pseudo-Euclidean plane, the equation of geodesics (straight-lines) from a
point $P_0\equiv(x_0,\,y_0)$ can be obtained from differential geometry 
\cite{ca, ei}. So for 
$|x-x_0|>|y-y_0|$, a straight-line (of the first kind) can be written as:
\be
(x-x_0)\sinh \th -(y-y_0)\cosh\th=0;  \label{c0} 
\ee
while for $|x-x_0|<|y-y_0|$ a straight-line (of the second kind) can be 
written as\footnote{the angle $\th'$ is referred to $y$ axis as stated in 
tab. \ref{tab2}}:
\be
(x-x_0)\cosh \th'-(y-y_0)\sinh\th'=0. \label{c1}
\ee 
Then from the topological characteristics of the pseudo-Euclidean plane, it
follows that the straight-lines will have the two possible expressions given 
by eqs. (\ref{c0}, \ref{c1}). \\ The use of extended \hy functions would give
just one equation for all the straight-lines, but in this paragraph and in the 
next one, we use the  classical \hy trigonometric functions which make more 
evident the peculiar characteristics of the pseudo-Euclidean plane.  \\
\noindent {\it $\bullet$ Pseudo-orthogonality.} \\
As in the Euclidean plane, two straight-lines in the pseudo-Euclidean plane
are said pseudo-orthogonal when the scalar product of their versors is zero 
\cite{ei}. \\  It is easy to show that two straight-lines of the same kind,
as given by eqs. (\ref{c0}) or (\ref{c1}), can never be pseudo-orthogonal. 
Indeed a straight-line of the first kind (eq. \ref{c0}) has a 
pseudo-orthogonal line of the second kind (eq. \ref{c1}) and with the same 
angle ($\th=\th'$), and conversely\footnote{It is known that in the
complex formalism, the equation of a straight-line is given by 
$\Re \{ (x+iy)(\exp [i\f] )+A+iB=0 \}$, and its orthogonal line by
$\Im \{(x+iy)(\exp [i\f])+A+iB=0\}$. In the pseudo-Euclidean plane in the \hy
formalism the equation of a straight-line is
$\Re \{ (x+hy)(\exp [h\th] )+A+hB=0 \}$ and the hyperbolic part $\Im \{ (x+hy)(\exp [h\th])
+A+hB=0 \}$ is its pseudo-orthogonal line. \\
Note that the product of the angular coefficients for two pseudo-orthogonal 
lines is +1.}. Then, as it is well known in special
relativity  \cite[p. 479]{ef}, \cite{Na}, {\it two straight-lines are pseudo-orthogonal
if they are symmetric with respect to a couple of lines parallel to the axes 
bisectors} (fig. \ref{rettepo}).              \\
\noindent {\it $\bullet$ Axis of a segment.} \\
Let us consider two points  $P_1(x_1,y_1)$, $P_2(x_2,y_2)$. The points that 
have the same pseudo-Euclidean distance from these two points are determined
by the equation:
$$ \overline{PP_1}^2 = \overline{PP_2}^2 \Ra
(x-x_1)^2 - (y-y_1)^2 = (x-x_2)^2 -(y-y_2)^2.
$$
This implies that :
\be
(x_1-x_2)(2x-x_1-x_2)=(y_1-y_2)(2y-y_1-y_2)\;,
\label{AssePapa}
\ee
and in canonical form:
\be
y = {(x_1 - x_2) \over (y_1 - y_2)}\;x +
{ {(y_1^2 - y_2^2) - (x_1^2 - x_2^2)} \over {2 (y_1 - y_2)}}
\label{Asse}
\ee
eqs. (\ref{AssePapa}) and (\ref{Asse}) show that the axis of a segment in 
the pseudo-Euclidean plane has the same properties as in the Euclidean plane:
i.e., it is pseudo-orthogonal to the segment $P_1P_2$ in its middle point.

\noindent {\it $\bullet$ Distance of a point from a straight-line.} \\
Let us take a point $P(x,\,y)$ on a straight-line of the second kind 
$\g :\{ y- mx - q=0 ;\; |m|>1\}$, and a point  $P_1(x_1,y_1)$ outside the 
straight-line. The "square distance"
$\overline{P\,P}_1^2 =(x-x_1)^2-(y-y_1)^2$ has its extreme for
 $x\equiv x_2 = (x_1- m y_1- mq)/(1- m^2)$, with a square distance:
\be
\overline{P_2P_1}^{2}\equiv D_{1,\, 2}= {{ (y_1 -mx_1-q)^2} \over {m^2-1}} 
\mbox{ and } d_{1,\,2} = {{ |y_1 -mx_1-q|} \over {\sqrt{|m^2-1|}}}\,. \\. 
\label{PuntoRetta}
\ee
It is easy to control that this distance correspond to a {\it maximum } as it 
is well known from special relativity \cite{Na}, \cite[p. 315]{Grav}. \\
From the expression (\ref{PuntoRetta}) it follows that: {\it the linear 
distance from the point $P_1$ to the straight line $\c$, is proportional to 
the result of substituting the coordinates of $P_1$ in the equation of $\c$,
as well as in Euclidean geometry. } \\
The equation of the straight-line $P_1,\,P_2$ is:
\[
(y-y_1) = {1 \over m} (x-x_1)
\]
that represents a straight-line {\it pseudo-orthogonal} to $\g$.\\
\section{Trigonometry in the pseudo-Euclidean plane}
\subsection{Goniometry}
By using the \hy  Euler formula (eq. \ref{cosi}) we can derive the \hy  
angles addition formulas \cite{Fj, Fj2}:
\be
\cosh_e[\a\pm\b]+h\sinh_e[\a\pm\b]=(\cosh_e[\a]+h\sinh_e[\a])(\cosh_e[\b]
\pm h\sinh_e[\b]). \label{somma}
\ee
These formulas allow to obtain for \hy  trigonometric functions all the expressions that are 
equivalent to the Euclidean goniometry ones.
\subsection{Trigonometry} \label{trigpe}
Let us consider a triangle in the pseudo-Euclidean plane with no sides 
parallel to axes bisectors: let us call $P_n\equiv(x_n,\,y_n)\; n=i,\,j,\,k
\,|\; i\neq j\neq k$ the vertices, $\th_n$ the \hy  angles. 
The square \hy  length of the side opposite to vertex $P_i$ is defined
by eq. (\ref{disq1}):
\be
D_i\equiv D_{j,\, k}=(z_j-z_k)(\tilde z_j-\tilde z_k) \mbox{ and } d_i=\sqrt{|D_i|}. 
\label{disq3}
\ee
as pointed out before $D_i$ must be taken with its sign. \\
Following the conventions of the Euclidean trigonometry we associate to the sides
three vectors oriented from $P_1\ra P_2;\;P_1\ra P_3 ;\;P_2\ra P_3$. \\
From (\ref{cos}), (\ref{sen}) and taking into account the sides orientation as done 
in the Euclidean trigonometry, we obtain:
\ba
\cosh_e \th_1 =& {\displaystyle\frac{(x_2-x_1)(x_3-x_1)-(y_2-y_1)(y_3-y_1)}{d_2\,d_3}}; \nonumber\\ 
\sinh_e \th_1 =& {\displaystyle\frac{(x_2-x_1)(y_3-y_1)-(y_2-y_1)(x_3-x_1)}{d_2\,d_3}}; \nonumber\\
\cosh_e \th_2 = &-{\displaystyle\frac{(x_3-x_2)(x_2-x_1)-(y_3-y_2)(y_2-y_1)}{d_1\,d_2}};\nonumber\\ 
\sinh_e \th_2 = &{\displaystyle\frac{(x_2-x_1)(y_3-y_2)-(y_2-y_1)(x_3-x_2)}{d_1\,d_2}}; \nonumber\\
\cosh_e \th_3 = &{\displaystyle\frac{(x_3-x_2)(x_3-x_1)-(y_3-y_2)(y_3-y_1)}{d_1\,d_3}};\nonumber\\ 
\sinh_e \th_3 = &{\displaystyle\frac{(x_3-x_1)(y_3-y_2)-(y_3-y_1)(x_3-x_2)}{d_1\,d_3}}. \;\;
\label{ele} 
\ea 
It is straightforward  to verify 
that all the functions $\sinh_e\th_n$ have the same numerator. If we call this
numerator: 
\be
x_1(y_2-y_3)+x_2(y_3-y_1)+x_3(y_1-y_2)=2S \label{Sup} 
\ee
we can write:
\be
2S=d_2\,d_3\sinh_e\th_1=
d_1\,d_3\sinh_e\th_2=d_1\,d_2\sinh_e\th_3 \label{area}.
\ee
In the Euclidean geometry a quantity equivalent to $S$ represents the  
area of the triangle.
In the pseudo-Euclidean geometry $S$ is still {\it an
invariant quantity linked to the triangle}. For this reason it is
appropriate to call $S$ the {\it pseudo-Euclidean area} \cite{Ya2}. We note that
the expression of area (eq. \ref{Sup}), in terms of vertices coordinates, is
exactly the same as in the Euclidean geometry (Gauss formula for a polygon area
applied to a triangle). \\
Now we are able to restate the trigonometric laws in the
pseudo-Euclidean plane.\\
\noindent {\it $\bullet$ Law of sines.}\\
{\it In a triangle the ratio of the \hy  sine to the \hy  length of the 
opposite side is constant}:
\be
\frac{\sinh_e\th_1}{d_1}=\frac{\sinh_e\th_2}{d_2}=
\frac{\sinh_e\th_3}{d_3}. \label{tsen}
\ee
{\it Proof.}
This theorem follows from eq. (\ref{area}) if we divide it by $d_1\,d_2\,d_3$.

Then {\it if two \hy triangles have the same \hy angles, they will have their
sides proportional.}\\
\noindent {\it $\bullet$ Laws of cosines.}\\
From the definitions of the side lengths (eq. \ref{mod}) and \hy angular
functions given by eqs. (\ref{ele}) it is easy to verify that\footnote{
Obtained by another method in \cite{Fj2}.}:
\be
D_i=D_j+D_k-2d_j\,d_k\cosh_e\th_i,  \label{carnot}
\ee
and:
\be
d_i=|d_j\cosh_e \th_k+d_k\cosh_e \th_j|.  \label{proiezioni}
\ee
\noindent {\it $\bullet$ Pythagoras theorem.}\\     
Let us take in eq. (\ref{carnot}) $\cosh_e\th_i=0$. From eq. (\ref{ele}) it 
follows: \\
\be
(x_j-x_i)(x_k-x_i)=(y_j-y_i)(y_k-y_i)\,\,\Ra\,\,\frac{y_j-y_i}{x_j-x_i}\equiv
m_{ij}=\frac{1}{m_{ik}}\equiv\frac{x_k-x_i}{y_k-y_i}.
\ee
The side $P_iP_j$ is pseudo-orthogonal to the side $P_iP_k$, i.e. the triangle is a 
right triangle and we obtain the hyperbolic Pythagoras theorem:
\be
D_i=D_j+D_k.   \label{pitagora}
\ee

We have seen that the topology of the pseudo-Euclidean plane is more complex 
with respect to the Euclidean one, as well as the relations between $\sinh_e$ 
and $\cosh_e$ and between the side lengths and the square lengths of the sides. 
For these reasons we could think that the triangle solution would require
more information, nevertheless the results of appendix (\ref{soltri}) show that as in
the Euclidean trigonometry: {\it all the sides and angles of a triangle can
be determined if we know three elements (with at least one side)}. \\
It also follows the \\ {\sc Theorem}
{\it all the elements of a triangle are invariant for \hy  rotation}.\\
{\it Proof. }
Let us consider a triangle with vertices in the points
\be
P_1\equiv (0,\,0), P_2\equiv (x_2,\,0) \mbox{ and }P_3\equiv
(x_3,\,y_3): \label{tripar} 
\ee
since $\overline{P_1P}_3\equiv d_2$  and $\overline{P_1P}_2\equiv d_3$ are 
invariant quantities, from eq.
(\ref{invpol}) follows that $\th_1$ is invariant too. Since these
three elements determine all the others, all the elements will be invariant.

From the exposed invariance it follows that, by the coordinate axes translation and
\hy  rotation, any triangle can be put with a vertex in $P\equiv(0,\,0)$, and
a side on one coordinate axis. \\
Then we do not loss in generality if, from now on, we consider a triangle in a 
position that will facilitate the control of the theorems which follow.  
Consequently we will consider the triangle with vertices in the points given 
by eq. (\ref{tripar}) and \hy  square lengths:
\be
D_1 = (x_3-x_2)^2-y_3^2\,;\hspace{8mm}
D_2 = x_3^2-y_3^2\,;\hspace{8mm}
D_3 = x_2^2\,.
\ee
By using eqs. (\ref{mod}), (\ref{ele}) we obtain the other elements:
\be      \label{trior}
\begin{tabular}{lll}
$\cosh_e\th_1={\displaystyle\frac{x_2\,x_3}{d_2d_3}}$; & \vspace{2ex}
$\cosh_e\th_2={\displaystyle\frac{x_2(x_2-x_3)}{d_1d_3}}$; &
$\cosh_e\th_3={\displaystyle\frac{x_3(x_3-x_2)-y_3^2}{d_1d_2}}$ \\
$\sinh_e\th_1={\displaystyle\frac{x_2\,y_3}{d_2d_3}}$;&
$\sinh_e\th_2={\displaystyle\frac{x_2\,y_3}{d_1d_3}}$;&
$\sinh_e\th_3={\displaystyle\frac{x_2\,y_3}{d_1d_2}}$
\end{tabular}
\ee
\subsection{The triangle angles sum} \label{sommaangoli} 
In an Euclidean triangle given two angles ($\f_1,\,\,\f_2$), the third one
($\f_3$) can be found using the relation $\f_1+\f_2+\f_3=\pi$. This relation
can be expressed in the following form:
$$\sin(\f_1+\f_2+\f_3)=0\,,\hspace{8mm} \cos(\f_1+\f_2+\f_3)=-1\,$$
 that allows a direct control for a pseudo-Euclidean triangle.

In fact by eq. (\ref{somma}) and using the relations (\ref{trior}) we obtain:
\small
$$\sinh_e(\th_1+\th_2+\th_3)\equiv\sinh_e\th_1\sinh_e\th_2\sinh_e\th_3+$$ $$+\sinh_e\th_1\cosh_e\th_2\cosh_e\th_3
+\cosh_e\th_1\sinh_e\th_2\cosh_e\th_3+\cosh_e\th_1\cosh_e\th_2\sinh_e\th_3\equiv $$
$$\frac{x_2^2y_3[x_2\,y_3^2+x_2\,x_3(x_2-x_3)+x_3^2(x_3-x_2)
-x_3\,y_3^2-x_3(x_2-x_3)^2-y_3^2(x_2-x_3)]}{d_1^2 d_2^2 d_3^2}=0,\vspace{2.5ex}$$
$$\cosh_e(\th_1+\th_2+\th_3)\equiv\cosh_e\th_1\cosh_e\th_2\cosh_e\th_3+\sinh_e\th_1\sinh_e\th_2\cosh_e\th_3+
$$ $$+\sinh_e\th_1\cosh_e\th_2\sinh_e\th_3+\cosh_e\th_1\sinh_e\th_2\sinh_e\th_3\equiv $$
$$\frac{x_2^2\{-x_3^2(x_2-x_3)^2+y_3^2[-x_3(x_2-x_3)+x_2\,x_3]+
y_3^2[x_2(x_2-x_3)-x_3(x_2-x_3)-y_3^2]\}}{d_1^2d_2^2 d_3^2}\equiv$$
\be
\equiv \frac{x_2^2\{-x_3^2[(x_2-x_3)^2-y_3^2]+y_3^2[(x_2-x_3)^2-y_3^2]\}}
{d_1^2d_2^2 d_3^2}\equiv -\frac{D_1 D_2 D_3}
{d_1^2d_2^2 d_3^2}=\mp 1. \label{coshsomma}
\ee
\normalsize
Then:
$\sinh_e(\th_1+\th_2+\th_3)=0,\,\,\cosh_e(\th_1+\th_2+\th_3)=\pm 1$.
By utilising eq. (\ref{thk}) and the formalism exposed in app. (\ref{estip}) we can say that 
{\it the sum of the triangle angles is given by:}
\be
(\th_1)_k+(\th_2)_{k'}+(\th_3)_{k''}\equiv (\th_1+\th_2+\th_3 )_{k\cdot
k'\cdot k''}=(0)_{\pm 1}                       \label{anglesum}
\ee
This result allows to state that if we know two angles we can
determine if the Klein group index of the third angle is
$\pm h: \{\th\in Us,\,Ds\}$ or $\pm 1: \{\th\in Rs,\,Ls\}$. From this we obtain 
the relation between $\cosh_e$ and $\sinh_e$ as stated by eqs. (\ref{quad}). 
This relation and the condition (\ref{anglesum}) allow to obtain the \hy  
functions of the third angle. Then: {\it also for
the pseudo-Euclidean triangles if we know two angles we can obtain the third one.}
In Appendix (\ref{soltri}) we show two examples of triangle solutions.
The triangle solutions for the other cases not considered there can be easily 
verified following the standard Euclidean approach.
\subsection{Equilateral hyperbolas in the pseudo-Euclidean plane} \label{iper}
The unit circle for the definition of trigonometric functions has its 
counterpart, in the pseudo-Euclidean plane, in the four arms of the unit
equilateral hyperbolas $|x^2-y^2|=1$, as shown in \cite{Fj}. 
Indeed the equilateral hyperbolas have many of the properties of circles
in the Euclidean plane; here we point out some of them. \\
{\it Definitions.}
If $A$ and $B$ are two points on the equilateral \hyp, the segment $AB$ is 
called a {\it chord} of the \hyp. We define two kinds of chords: if the points
$A,\,\,B$ are on the same arm of hyperbola we have {\it ``external chords"},
if the points are in opposite arms we have {\it ``internal chords"}.
Any internal chord that passes through the centre
``$O$" of the hyperbola  is called a {\it diameter of the \hyp}. We will call
$p$ the semi-diameter. \\
In particular we follow the definitions stated for 
segments and straight lines and call {\it hyperbolas of the first (second) kind 
if the tangent straight lines are of the first (second) kind. } 
If we introduce the square semi-diameter $P$ with its sign, and 
$p=\sqrt{|P|}$, we have $P<0$ ($P>0$) for hyperbolas of the first (second) kind.
\\ The following theorems hold\footnote{Some of these theorems are reported,
without demonstrations, in \cite{Ya2}}: \\
{\sc Theorem 1. }
{\it the diameters of the hyperbola  are the internal chords of minimum length}.\\
{\it Proof. }
We do not loss in generality if we consider \hyp s of the second kind, with 
their centre in the coordinate axes origin; then we have $A\equiv(p\cosh\th_1, 
p\sinh\th_1)$,
$B\equiv(-p\cosh\th_2, -p\sinh\th_2)$. The square length of the chord
is $\overline{AB}^2=p^2[(\cosh\th_1+\cosh\th_2)^2-(\sinh\th_1+\sinh\th_2)^2]
\equiv 4p^2\cosh^2[(\th_1-\th_2)/2]$, i.e., $\overline{AB}^2=4p^2$ for 
$\th_1=\th_2$ and $\overline{AB}^2> 4p^2$ for $\th_1\neq\th_2$.

Now we enunciate for equilateral hyperbolas the pseudo-Euclidean counterpart
of well-known Euclidean theorems for circles. These theorems can be 
demonstrated by elementary analytic geometry, as the previous theorem 1. \\   
{\sc Theorem 2. } {\it The line joining $O$ to the midpoint $M$ of a 
chord is pseudo-orthogonal to it.}  \\
This theorem is also valid in the limiting position when the chord
becomes tangent to the \hyp:       \\ 
{\sc Theorem 3. } {\it If $M$ is on the \hyp, the tangent at $M$ 
is pseudo-orthogonal to the diameter $OM$.} \\
{\sc Theorem 4. }
{\it If we have the points $A$ and $B$ on the same arm of the \hyp, for any 
point $P$ between $A$ and $B$, the hyperbolic angle $\widehat{APB}$ is half 
the hyperbolic angle $\widehat{AOB}$.} \\
{\sc Theorem 5. }
If $Q$ is a second point between $A$ and $B$, then $\widehat{APB}=
\widehat{AQB}$.\\
{\sc Theorem 6. }
{\it If a side of a triangle inscribed in an equilateral hyperbola passes
through the centre of the hyperbola, the other two sides are pseudo-orthogonal}.\\
{\sc Theorem 7. }
{\it  If we have three non-aligned points that can be considered the vertices 
of a triangle, there is always an equilateral hyperbola ({\sf circumscribed
\hyp)} which passes by these points.}\\
For a circumscribed hyperbola it is quite straightforward to obtain:
\be
P=-\frac{D_1 D_2 D_3}{4[(x_1-x_2)(y_1-y_3)-(y_1-y_2)(x_1-x_3)]^2}
\equiv -\frac{D_1 D_2 D_3}{16\,S^2}     \label{p3}
\ee
for $P>0$ we have an equilateral hyperbola of the second kind,  while
for $P<0$ we have an equilateral hyperbola of the first kind.
Then in relation to the hyperbola type we could say that there are two kinds
of triangles depending on the sign of $D_1 D_2 D_3$. 
From eqs. (\ref{p3}) and (\ref{area}) we obtain:
\be
p=\frac{d_1d_2 d_3}{4\,S}\equiv \frac{d_n}{2\sinh_e\th_n}
\ee
that is the same relation as that for the radius of a circumcircle
in an Euclidean triangle. \\
The theorems shown above indicate that in some cases solutions of equilateral 
hyperbolas problems may be more easily found by applying the exposed theory.
\section{Conclusions}
In this paper we have shown that the invariant quantities with respect to special
relativity Lorentz group allow an operative definition of \hy  trigonometric
 functions. These definitions together with the formal substitution of the
pseudo-Euclidean distance and \hy trigonometric functions respectively to
Pythagorean distance and circular trigonometric functions, allow to extend 
the classic trigonometry theorems to the Pseudo-Euclidean plane. From these theorems
it follows that the pseudo-Euclidean trigonometry can be  formalised with the same
coherence as the Euclidean trigonometry does and the \hy triangles can be solved  exactly
in the same way as the Euclidean triangles.
This can be done in spite of the more complex topology of the pseudo-Euclidean
plane with respect to the Euclidean plane. \\
This formalisation of the pseudo-Euclidean trigonometry
allows also to observe in a more general way \cite{Ya2} the Euclidean 
geometry and might contribute to give a concrete mathematical meaning        
to the pseudo-Euclidean plane geometry. 
As a final conclusion note that our results have been obtained by using the 
group properties of \hy numbers and by comparing hyperbolic and complex
numbers. Moreover as briefly resumed in note (\ref{hy}) and diffusely
showed in \cite{Ca} all systems of commutative hypercomplex numbers can be 
associated to finite and infinite Lie groups. These properties make 
these number systems suitable to be used in applied sciences. Unfortunately, 
none of the geometries associated with number systems of more than two 
units corresponds to the multidimensional Euclidean geometry or to four
dimensional space-time geometry used to describe the physical world. 
Then, the possibility of applying multidimensional commutative hypercomplex 
numbers to multidimensional geometries would require some radically new ideas. 
On the contrary the properties of \hy numbers
may allow a more complete insight into the pseudo-Euclidean geometry
just by using the formal correspondence between the complex and \hy numbers.
\section{Acknowledgments}
We wish to thank Prof. Dino Boccaletti for his suggestions to extend the
complex numbers applications to the hypercomplex numbers systems and for the
fruitful discussions during the development of our work.
\appendix
\section{Rotation invariants in the Euclidean plane} \label{inve}
Euclidean geometry studies the figure properties that do not depend
on their position in a plane. If these figures are represented in a Cartesian 
plane we can say, in group language, that Euclidean geometry studies the 
invariant properties by coordinate axes roto-translations.
It is well known that these properties can be expressed by complex numbers. 
Let us consider the Gauss-Argand complex plane where a 
vector is represented by $v=x+i\,y$. The axes rotation of an angle $\a$ 
transforms this vector in the new vector $v'\equiv v\exp [i\a]$. 
Therefore we can promptly verify that the quantity (as it is usually
done we call $\bar v=x-i\,y$)
$|v'|^2 \equiv v'\bar {v'}= {v\exp [i\a]\bar v\exp [-i\a]}
\equiv |v|^2$ is invariant by axes rotation.
In a similar way we find two invariants related to any couple of vectors. If 
we consider two vectors: $v_1=x_1+iy_1$, $v_2=x_2+iy_2$; we have that
{\it the real and the imaginary part of the product $v_2 \bar v_1$ are 
invariant by axis rotation.}  In fact $v^{'}_2\bar v_1^{'}= v_2\exp
[i\a]\bar v_1\exp [-i\a]\equiv v_2\,\bar{v}_1$. 
Now we will see that these two invariants allow an operative definition of 
trigonometric functions. Let us represent the two vectors in polar coordinates: 
$v_1\equiv \r_1\exp[i\f_1],\,v_2\equiv \r_2\exp[i\f_2]$. Consequently we have: 
\be
v_2\bar v_1=\r_1\r_2\exp[i(\f_2-\f_1)]\equiv
\r_1\r_2[\cos(\f_2-\f_1)+i\sin(\f_2-\f_1)]. \label{polc}
\ee
As it is well known the resulting real part of the vector product represents 
the scalar product, while the imaginary part represents the module of the
vector product, i.e., the area of the parallelogram defined by the two vectors. \\
The two invariants of eq. (\ref{polc}) allow an operative 
definition of  trigonometric functions.
In Cartesian coordinates we have: 
\be
v_2\,\bar v_1=(x_2+i\,y_2) (x_1-i\,y_1)\equiv x_1\,x_2+y_1\,y_2+i(x_1\,y_2-x_2\,y_1),
\label{carc}
\ee
and by using eqs. (\ref{polc}) and (\ref{carc}) we obtain:
\be
\cos(\f_2-\f_1)=\frac {x_1\,x_2+y_1\,y_2}{\r_1
\,\r_2};\;\;\;\;\;\;\;\;\;\sin(\f_2-\f_1)=\frac{x_1\,y_2-x_2\,y_1}{\r_1\,\r_2}
\label{cir}
\ee
We know that the theorems of Euclidean trigonometry are usually obtained by 
following a geometric approach. Now by using the Cartesian representation of 
trigonometric functions, given by eqs. (\ref{cir}), it is straightforward to 
control that the
trigonometry theorems are simple identities. In fact let us define a triangle
in a Cartesian plane by its vertices $P_n\equiv(x_n,\,y_n)$: from the
coordinates of these point we obtain the side lengths and, from the eqs. 
(\ref{cir}) the angles trigonometric functions. By these definitions it is 
easy to control the identities defined by the trigonometry theorems.
\section{Hyperbolic  functions on unit equilateral hyperbolas}
\label{estip}
In the complex Gauss-Argand plane the goniometric circle used for the 
definition of trigonometric functions is expressed by $x+i\,y=\exp[i\f]$. 
In the \hy  plane the \hy  trigonometric functions can be defined on the
unit equilateral hyperbola, that can be expressed in a way similar to
goniometric circle: $x+h\,y=\exp[h\th]$. However this expression represents 
only the arm of unit equilateral hyperbolas in the {\it Right sector }($Rs$).
If we want to extend the \hy functions on the whole plane, we must take into
account all the unit equilateral hyperbola arms, given by 
$x+h\,y=\pm \exp[h\th]$ and $x+h\,y=\pm h\,\exp[h\th]$ for which 
$|x^2-y^2|=1$. \\ 
The aim of this appendix is to summarise the approach followed in \cite{Fj} 
to demonstrate how these unit
curves allow to extend the \hy functions and to obtain the addition formula
for angles in any sector. \\
These unit curves are the set of points $U$, where $U=\{z\,|\,\r(z)=1\}$. 
Clearly $U(\cdot)$ is a group, subgroups of this group are for $z\in
Rs$, and for $z\in Rs+Us$, $z\in Rs+Ls$, $z\in Rs+Ds$.
For $z\in Rs$ the group $U(\cdot)$ is isomorphic to $\th(+)$ where $\th\in 
{\mathbf R}$ is the angular function that for $-\infty<\th<\infty$ traverses 
the $Rs$ arm of the unit hyperbolas. Now we can have a complete isomorphism 
between
$U(\cdot)$ and the angular function, extending the last one to other sectors.
This can be done by the Klein four group: $k\in K=\{1,\,h,\,-1,\,-h\}$.\\ 
Indeed let us consider the expressions of the four arms of the hyperbolas
(tab. \ref{hip} for $\r=1$). We can extend the angular functions as a product between
$\exp[h\,\th]$ and the Klein group: $k\cdot \exp[h\,\th]$. We can write:
$U=\{k\exp[h\,\th]\,|\, \th\in {\mathbf R},\, k\in K\}$, and we call 
$U_k=\{k\exp[h\,\th]\,|\, \th\in{\mathbf R}\}$, the hyperbola arm with the value $k$.
In the same way we call $\th_k$ the ordered pair $(\th, k)$, and we define
$\Th\equiv{\mathbf R}\times K=\{\th_k\,|\, \th\in{\mathbf R},\,\, k\in K\}$ and 
$\Th_k\equiv{\mathbf R}\times \{k\}=\{\th_k\,|\, \th\in{\mathbf R}\}$.
$\Th_1$ is isomorphic to ${\mathbf R}(+)$; then we accordingly think of $\Th(+)$ as an 
extension of ${\mathbf R}(+)$. To define the complete isomorphism between $\Th(+)$
and $U(\cdot)$, we have to define the addition rule for angles 
$\th_k+\th_{k'}'$. This rule is obtained from the isomorphism itself:\\
\be
\th_k+\th_{k'}'\Ra U_k\cdot U_{k'}\equiv k\exp[h\th]\cdot
k'exp[h\th']\equiv k\,k'exp[h(\th+\th')]\Ra (\th+\th')_{k\,k'}. \label{thk}
\ee
On these basis it is straightforward to obtain the hyperbolic angle $\th$ and 
the Klein index ($k$) from the extended hyperbolic functions 
$\sinh_e\th$ and $\cosh_e\th$:
\ba
\mbox{If } |\sinh_e\th|<|\cosh_e\th| \Ra &\th=\mbox{tanh}^{-1}\left(\frac{
\sinh_e\th}{\cosh_e\th}\right);\;\;\;& k=\frac{\cosh_e\th}{|\cosh_e\th|}\cdot 1 \nonumber \\
\mbox{If } |\sinh_e\th|>|\cosh_e\th| \Ra &\th=\tanh^{-1}\left(\frac{
\cosh_e\th}{\sinh_e\th}\right);\;\;\;& k=\frac{\sinh_e\th}{|\sinh_e\th|}\cdot h. \nonumber
\ea
\section{Two examples of \hy triangle solutions} \label{soltri}
In this appendix we report two elementary examples of \hy  triangle solution
in order to point out analogies and differences with the Euclidean
trigonometry. We will note that the Cartesian representation can give some 
simplifications in the triangle solution. The Cartesian axes will be chosen 
so that $P_1\equiv(0,\,0)$, $P_2\equiv(\pm d_3,\,0)$, or $P_2\equiv(0,\,\pm d_3)$. The two possibilities for
$P_2$ depend on the sign of $D_3$, the sign plus or minus is chosen
so that one goes from $P_1$ to $ P_2$ to $P_3$ counter-clockwise. Thanks to relations 
(\ref{trior}), the triangle is completely determined by the $P_3$ coordinates. 

\noindent {\it $\bullet$ Given two sides and one opposite angle}\\
$\th_1;\;D_1;\;D_3$ with $D_3>0$. \\ 
Applying the first law of cosine to the side $d_1$ we have: \\
$D_2-2d_2d_3\cosh_e\th_1+D_3-D_1=0$, from which we can obtain
$d_2$. In fact:\\
for $\cosh_e\th_1>\sinh_e\th_1$
\[
D_2=d_2^2\Ra \,
d_2=d_3\cosh_e\th_1\pm\sqrt{d_3^2\sinh_e^2\th_1+D_1}\,,
\]
for $\cosh_e\th_1<\sinh_e\th_1$
\be
D_2=-d_2^2\Ra \,
d_2=-d_3\cosh_e\th_1\pm\sqrt{d_3^2\sinh_e^2\th_1-D_1}\,. \label{duelati}
\ee
\normalsize
So, as for the Euclidean counterpart, we can have, depending on the value of
the square root argument, two, one or no solutions. \\
Now the vertex $P_3$ coordinates are given by:
\be
x_3=d_2\cosh_e\th_1 ;\;\; y_3=d_2\sinh_e\th_1.
\label{tria1}
\ee
We now use an analytic method distinctive of Cartesian plane.
The coordinates of the point $P_3$ can be obtained intersecting the 
straight-line $y=\tanh_e\th_1\,x$ with the hyperbola centred in $P_2$ and 
having square semi-diameter
$P=D_1$, i.e., by solving the system:
\be
y=\tanh_e\th_1\,x;\;\;\;\;\;\;(x-d_3)^2-y^2=D_1.             \label{tria4}
\ee
The results are the same as those of eq. (\ref{duelati}), but now it is easy
to understand the geometrical meaning of the solutions.
In fact if $D_1>0$ and  $d_1>d_3$ the point $P_1$ is included in an 
hyperbola arm and we will have always two solutions. Otherwise if
$\sinh\th_1<d_1/d_3$ there are no solutions, if $\sinh\th_1=d_1/d_3$ there
is just one solution and if $\sinh\th_1>d_1/d_3$ two solutions. \\
If $D_3<0$ the $P_2$ vertex must be put on the $y$ axis and we will have 
the system: 
$$x=\tanh_e\th_1\,y;\;\;\;\;\;\;(y-d_3)^2-x^2=-D_1.$$ By comparing this 
result with the solutions 
of the system (\ref{tria4}) we have to change $x\lra y$. 

\noindent {\it Given two angles and the side between them}\\
$\th_1;\;\th_2;\;D_3$ and $D_3>0$.\\ In the Euclidean  trigonometry the
solution of this problem is obtained using the condition that the sum of the 
triangle angles is $\pi$.
We will use this method, that will allow to familiarise with the Klein 
group defined in appendix (\ref{estip}), as well as an analytical geometry
method. Let us start with the classical method. We have: \\
$\th_3=-(\th_1+\th_2)$ and a Klein group index 
$k_3$ so that $k_1k_2k_3=\pm 1$. Then \hy  trigonometric functions will be
given by:  $\sinh_e \th_3=\sinh|\th_1+\th_2|$ if $k_3=\pm 1$ and by
$\sinh_e \th_3=\cosh(\th_1+\th_2) $ if $k_3=\pm h$.
From the law of sines we obtain: $d_2=d_3\frac{\sinh_e\th_2}{\sinh_e\th_3}.$
Eqs. (\ref{tria1}) allow to obtain the $P_3$  coordinates. \\
In the Cartesian representation we can use the following method:
let us consider the straight-lines
$P_1P_3\Ra y=\tanh_e\th_1\,x$ and $P_2P_3\Ra y=-\tanh_e\th_2\,(x-x_2)$. \\
By solving the system between these straight-lines we obtain the $P_3$ 
coordinates:
\be
P_3\equiv\left(x_2\frac{\tanh_e\th_2}{\tanh_e\th_2+\tanh_e\th_1},   \;
x_2\frac{\tanh_e\th_1\,\tanh_e\th_2 }{\tanh_e\th_2+\tanh_e\th_1}  \right).
\ee
If $D_3<0$ the straight-lines equations are:\\ $y=\coth_e\th_1\,x$ and $
\,y-y_2=-\coth_e\th_2\,x$. and the solution will be:
\be
P_3\equiv\left(y_2\frac{ \tanh_e\th_1 \tanh_e\th_2}{\tanh_e\th_2+\tanh_e\th_1},   \;
y_2\frac{\,\tanh_e\th_2 }{\tanh_e\th_2+\tanh_e\th_1}  \right).
\ee

\end{document}